# Hybrid 2D/3D photonic integration for non-planar circuit topologies

Aleksandar Nesic,[1,*] Matthias Blaicher,[1,2] Tobias Hoose,[1,2] Andreas Hofmann,[3] Matthias Lauermann,[1,4] Yasar Kutuvantavida,[1,2] Martin Nöllenburg,[5] Sebastian Randel,[1] Wolfgang Freude,[1] and Christian Koos,[1,2,4,**]

[1]*Karlsruhe Institute of Technology (KIT), Institute of Photonics and Quantum Electronics (IPQ), Engesserstrasse 5, 76131, Karlsruhe, Germany*
[2]*Karlsruhe Institute of Technology (KIT), Institute of Microstructure Technology (IMT), Hermann-von-Helmholtz-Platz 1, 76344 Eggenstein-Leopoldshafen, Germany*
[3]*Karlsruhe Institute of Technology (KIT), Institute for Automation and Applied Informatics (IAI), Hermann-von-Helmholtz-Platz 1, 76344 Eggenstein-Leopoldshafen, Germany*
[4]*Vanguard Photonics GmbH, Gablonzer Strasse 10, 76185 Karlsruhe, Germany*
[5]*TU Wien, Institute of Logic and Computation, Algorithms and Complexity Group, Favoritenstrasse 9-11, 1040 Vienna, Austria*
[*]aleksandar.nesic@kit.edu, [**]christian.koos@kit.edu

**Abstract:** Complex photonic integrated circuits (PIC) may have strongly non-planar topologies that require waveguide crossings (WGX) when realized in single-layer integration platforms. The number of WGX increases rapidly with the complexity of the circuit, in particular when it comes to highly interconnected optical switch topologies. Here, we present a concept for WGX-free PIC that rely on 3D-printed freeform waveguide overpasses (WOP). We experimentally demonstrate the viability of our approach using the example of a $4 \times 4$ switch-and-select (SAS) circuit realized on the silicon photonic platform. We further present a comprehensive graph-theoretical analysis of different $n \times n$ SAS circuit topologies. We find that for increasing port counts $n$ of the SAS circuit, the number of WGX increases with $n^4$, whereas the number of WOP increases only in proportion to $n^2$.

## 1. Introduction

Photonic integrated circuits (PIC) are becoming increasingly complex, incorporating thousands of photonic devices on a single chip [1,2]. The silicon photonic (SiP) platform, in particular, stands out to high integration density and offers high-yield fabrication on large-area substrates using mature CMOS processes [3,4]. However, as the complexity of PIC increases, non-planar circuit topologies with hundreds or even thousands of waveguide crossings (WGX) are unavoidable, and the number of WGX often increases in a strongly nonlinear way with the complexity of the circuit. As a consequence, compact WGX have evolved into key building blocks, and substantial research effort has been dedicated to optimizing their performance. This has led to remarkably low insertion loss (IL) of 0.017 dB and crosstalk as small as $-55$ dB at $\lambda = 1550$ nm, demonstrated for partially etched multi-mode interference (MMI) structures that feature a relatively large footprint of approximately $30 \times 30$ µm$^2$ [2]. Fully etched MMI structures allow to reduce the footprint to, e.g., $9 \times 9$ µm$^2$, but losses and crosstalk increase to, e.g., 0.028 dB and $-37$ dB, respectively [5]. Arrays of WGX can be compactly realized by exploiting Bloch modes in multi-mode waveguides: For SiP structures fabricated by electron-beam lithography, values of IL = 0.019 dB and crosstalk of less than $-40$ dB per WGX were demonstrated for a $101 \times 101$ WGX array with a 3 µm waveguide pitch [6]. For optical lithography, the best reported values for Bloch mode WGX are IL = 0.04 dB and crosstalk less than $-35$ dB for a $1 \times 10$ array of crossings with a 3.25 µm waveguide pitch [7].

However, while these demonstrations are impressive, even IL of the order of a few hundredths of dB and crosstalk of the order of – 40 dB can have a substantial impact on the performance of large-scale PIC that may comprise tens of thousands of WGX. A prime example in this context are high-radix switches that rely on the so-called switch-and-select (SAS) architecture [8]. The SAS scheme offers low crosstalk and simple control but requires a complex and highly non-planar interconnect network that provides a dedicated waveguide from each input to each output port. In fact, finding a layout that gives the minimum number $\eta_{n,n}$ of WGX in an $n \times n$ SAS circuit, and generally in any circuit, is an NP-complete problem [9], and $\eta_{n,n}$ scales with $n^4/16$ according to a still unproven conjecture [10,11]. This leads to tens of thousands of WGX for $n = 32$ and to approximately one million WGX for $n = 64$. To illustrate the associated performance penalty by WGX crosstalk, let us consider an example of a waveguide WG000 crossing an array of 100 other waveguides WG001 … WG100. Let us also assume that each WGX features a crosstalk of – 40 dB, i.e., an amplitude coupling coefficient of $10^{-2}$, and that there is an equal power level in WG001 … WG100. Considering a worst-case scenario where all 100 crosstalk signals would add coherently in WG000, the overall amplitude of the crosstalk signal would reach the same order of magnitude as each of the signals in WG001 … WG100. Moreover, a few hundredths dB of IL per WGX would result in several dB of IL along WG000. This example illustrates that large-scale PIC with highly non-planar topologies may not rely on WGX that are realized in single-layer integration platforms.

To overcome the limitations of conventional WGX, multi-layer PIC have been proposed exploiting multiple stacked waveguide layers, realized from silicon [12,13], silicon nitride ($Si_3N_4$) [14,15] or as a combination of both waveguide technologies [16–20]. The deposition of the upper layers is typically done by chemical vapor deposition (CVD) and involves chemical-mechanical planarization (CMP) of intermediate $SiO_2$ cladding layers that separate the waveguide layers. While simple two-layer implementations offer decent performance [18,20], three-layer structures have been shown to greatly reduce inter-layer crosstalk while maintaining efficient interlayer coupling [13,19]. This allows to reduce the crosstalk to less than – 56 dB with remarkably low interlayer coupling losses of less than 0.15 dB from the bottom to the top layer using a pair of vertical directional couplers of approximately 190 µm length per side [19]. However, while this approach would offer utmost scalability and the ability to cross entire groups of waveguides, it introduces additional technological complexity associated with growth and patterning of the top waveguide layers and integration thereof into multi-layer metal back-end stacks.

In this paper we demonstrate a technically simple hybrid 2D/3D photonic integration as an alternative to multi-layer circuits for realizing non-planar circuit topologies. Our approach is based on 3D-printed freeform polymer structures [21], which we refer to as freeform optical waveguide overpasses (WOP). WOP are realized in-situ by direct laser writing using two-photon polymerization [22]. This technique has previously been used for fabrication of so-called photonic wire bonds that enable low-loss single-mode connections across chip boundaries [23–25]. WOP offer low crosstalk of less than – 75 dB and allow to bridge series of parallel waveguides, thereby replacing a multitude of WGX. We demonstrate the viability of our approach by realizing a 4 × 4 SAS circuit. Based on a graph-theoretical analysis, we estimate that the number of WOP needed to realize a WGX-free $n \times n$ SAS PIC scales in proportion to $n^2/2$. A 64 × 64 SAS circuit would hence require only approximately 2000 WOP as opposed to the estimated one million conventional WGX. Fabrication of WOP may be efficiently combined with 3D-printing for die-level packaging [23–25], and offers the opportunity to locally incorporate multi-layer elements into standard SiP circuits, fabricated through readily available foundry services.

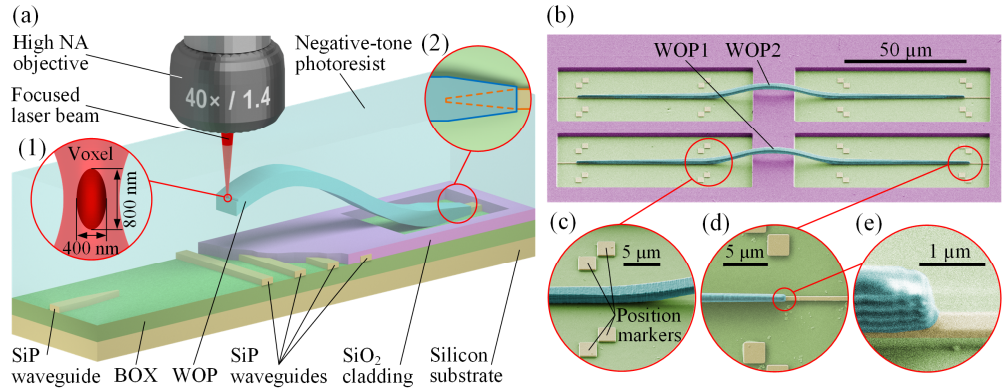

**Fig. 1.** Concept and implementation of waveguide overpasses (WOP) on the silicon photonic (SiP) platform. **(a)** The WOP is written into a liquid negative-tone photoresist that is deposited onto the PIC. For better coupling to the SiP on-chip waveguides, the $SiO_2$ cladding is locally removed down to the buried oxide (BOX) layer. Inset (1): The spatial resolution of the two-photon lithography is determined by the size of the volumetric pixel (voxel) that results from two-photon polymerization. Inset (2): Tapers in the WOP and in the SiP waveguide improve the coupling efficiency. **(b)** Scanning electron microscope (SEM) image of the WOP (colors were added by image processing). **(c)**–**(e)** Close-ups of different parts of the WOP. Position markers indicate the positions of the SiP waveguide ends that need to be interconnected. During fabrication of our chip, the $SiO_2$ cladding layer has been unintentionally over-etched, and part of the BOX has been unintentionally removed, see Subfigure (e).

The paper is structured as follows: In Section 2 we introduce the concept of 3D-printed WOP. A graph-theoretical analysis of the number of necessary WOP and WGX for realizing surface-coupled $n \times n$ SAS devices is provided in Section 3. Design and experimental testing of the demonstrator device are explained in Section 4. Appendix A provides definitions of graph theory terms. Appendix B gives further details of the graph-theoretical approach used for the analysis in Section 3. Appendix C gives a detailed graph-theoretical analysis of the number of necessary WOP and WGX for realizing facet-coupled SAS devices.

## 2. Concept of waveguide overpasses (WOP)

The concept of a 3D-printed freeform optical WOP is illustrated in Fig. 1 for the example of a SiP circuit. The PIC may be fabricated through standard processes offered by a commercial SiP foundry [26], including selective removal of $SiO_2$ cladding layer to access the tapers of the SiP waveguides that need to be interconnected. For fabrication of the WOP, a negative-tone photoresist is locally deposited onto the chip, and the WOP is then 3D-printed into the resist by direct laser writing based on two-photon polymerization. After exposure, the resist is removed, and the free-standing WOP structures are clad by a low-index polymer that acts as cladding and humidity protection (not shown in Fig. 1). Depending on the length, WOP may bridge tens or even hundreds of planar waveguides in the SiP device layer. Figure 1(b) displays scanning electron microscope (SEM) images of the two WOP on our demonstrator device before the cladding was applied, with colors added by image processing for better visualization. Figures 1(c)–(e) show close-ups of different parts of the lower WOP and demonstrate the accuracy of the direct laser writing method. The two-photon lithography system uses CMOS patterned silicon markers for automated detection of the SiP waveguides that need to be interconnected. The 3D-printing time of one WOP is about 30 s with a significant potential for further reduction. The refractive index of the WOP core material amounts to $n_{WOP} \approx 1.53$, and the cladding has a refractive index of $n_{cladding} \approx 1.36$ at 1550 nm. Note that the concept has been illustrated for the SiP platform here but can generally be applied to a wide range of PIC technologies.

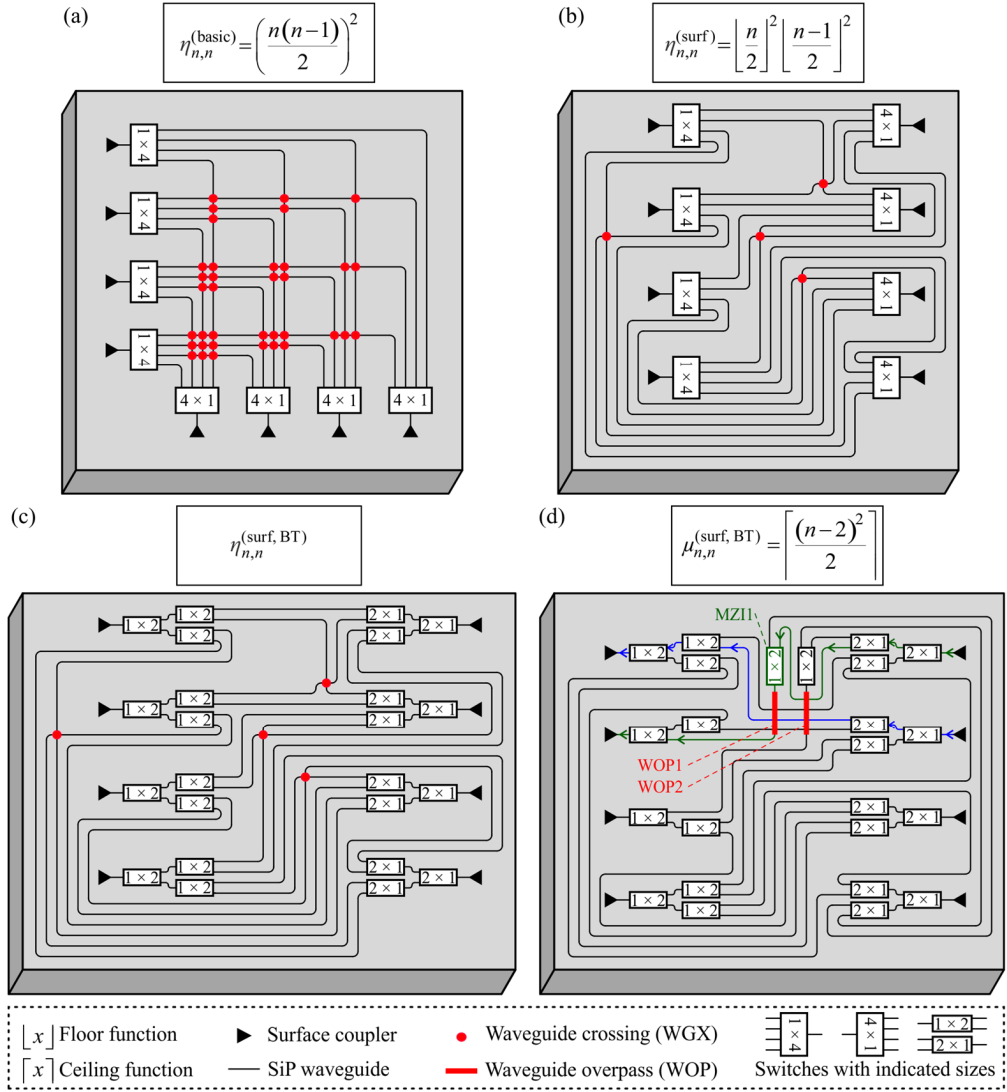

**Fig. 2.** Comparison of layouts of a 4 × 4 optical switch-and-select (SAS) circuit for surface coupling. **(a)** Basic layout for single-layer waveguide technology without any optimization for reduced numbers of waveguide crossings (WGX). **(b)** Optimal layout for single-layer waveguide technology, minimizing the number of WGX by routing of waveguides around the coupling elements. The formula for $\eta_{n,n}^{(\text{surf})}$ is a conjecture for the minimum possible number of WGX for an $n \times n$ SAS, if the $1 \times n$ and $n \times 1$ switches at the input and output ports are lumped elements (LE) [10,11]. For large port counts $n$, the number of WGX is conjectured to scale with $n^4/16$. **(c)** Best found, but not necessarily optimal layout for a single-layer 4 × 4 SAS circuit, in which the $1 \times 4$ and $4 \times 1$ switches have been realized as binary trees (BT) of $1 \times 2$ and $2 \times 1$ switches. A general analysis of this circuit topology for arbitrary $n$ is subject to ongoing investigations. **(d)** Best found, but not necessarily optimal WGX-free layout for hybrid 2D/3D circuits, minimizing the number of WOP. The switches are realized as BT in the same way as in (c). The formula for $\mu_{n,n}^{(\text{surf, BT})}$ is an upper bound for the minimum number of WOP. The optical paths that were used for the crosstalk measurement in Section 4 are marked in green (Path 1) and in blue (Path 2). The arrows indicate the direction of light propagation for the crosstalk measurement. The drive current of MZI1 is modulated by a sinusoidal signal for highly sensitive lock-in detection of the weak crosstalk signals.

## 3. Theoretical analysis of non-planar switch-and-select (SAS) circuit topologies

To experimentally demonstrate the viability of our approach, we use an $m \times n$ SAS circuit as an example of a PIC requiring many WGX. In the $m \times n$ SAS architecture, each of the $m$ input ports feeds a $1 \times n$ switch distributing the light to one of the $n$ output ports, and each of the $n$ output ports is fed by an $m \times 1$ switch, which selects light from one of the $m$ input ports. An illustration of a basic non-optimized implementation of a $4 \times 4$ SAS architecture is shown in Fig. 2(a), featuring a total number of 36 WGX in the depicted case, which would scale up to a total number of

$$\eta_{n,n}^{(\text{basic})} = \left(\frac{n(n-1)}{2}\right)^2 \tag{1}$$

for the case of an $n \times n$ SAS circuit. In the following, we show that these circuits can be realized with a significantly smaller number of WOP than the number of WGX, even if the layout of the circuit is optimized to reduce the number of WGX. To this end, we exploit graph theory to investigate the scaling of WGX and WOP number for increasing port counts $n$. For the remainder of this section, we consider the case where input and output ports are accessible from the top surface of the PIC and can hence be positioned anywhere on the chip. This case is referred to as *surface coupling*. Surface-coupled PIC may, e.g., rely on grating couplers, SiP waveguides that are bent upwards by ion implantation [27], or on 3D-printed lensed couplers [28]. We only give a summary of the results here; mathematical details can be found in Appendix B. In Appendix C, we also discuss the case of *facet coupling*, for which light is coupled to and from the PIC via waveguide facets along the chip boundary.

As a first step of the layout optimization, we exploit the fact that surface coupling allows to route waveguides around the couplers. This is illustrated in Fig. 2(b) for the example of a $4 \times 4$ SAS. In this implementation, we consider the $1 \times n$ and $m \times 1$ switches at the input and output ports as discrete entities that cannot be subdivided and that may hence be considered as lumped elements (LE). This leads to representation of the SAS circuit by a *complete bipartite graph* $K_{m,n}$ having two sets $M$ and $N$ of $m$ and $n$ *vertices*, respectively. Each vertex of set $M$ represents an input port of the SAS and its corresponding $1 \times n$ switch, and each vertex of set $N$ represents an output port and the associated $m \times 1$ switch. Each vertex of one set is connected to each vertex of the other set by a total of $mn$ *edges* that represent optical waveguides. In the following, we restrict our consideration to the particularly relevant cases of $K_{n,n}$, for which the number $m$ of input ports equals the number $n$ of output ports. A generalization to the case of $K_{m,n}$ can be found in Appendix B.

For conventional SAS implementations in single-layer waveguide technology, a layout with the smallest possible number of WGX can be achieved by optimizing the drawing of the corresponding graph model for finding the minimum number of *edge crossings* (or just *crossings*), which is an NP-complete problem [9]. Up to now [10], there is only a conjectured formula for the minimum possible number of crossings (*crossing number*), based on a straightforward graph drawing algorithm, only proven to give an upper bound [11],

$$\eta_{n,n}^{(\text{surf})} = \left\lfloor \frac{n}{2} \right\rfloor^2 \left\lfloor \frac{n-1}{2} \right\rfloor^2. \tag{2}$$

In this relation, $\lfloor x \rfloor$ denotes the floor function. For large $n$, the conjectured crossing number scales with $n^4/16$, thereby reducing the number of WGX by a factor of 4 compared to the simplistic non-optimized waveguide routing shown in Fig. 2(a). Note that the best published result for the lower bound of the crossing number in complete bipartite graphs $K_{n,n}$ states that for large $n$ the crossing number scales at least with $0.83 \cdot n^4/16 \approx n^4/19.28$ [29]. However, this is a theoretical result for the case of large $n$, which has not been supported by drawings of the corresponding graphs. In fact, for complete bipartite graphs, no drawings are known that lead to lower number of crossings than conjectured by Eq. (2). We therefore use the conjectured formula and its corresponding drawing as a basis for our analysis of the scaling of WGX for increasing port counts $n$. For an $n \times n$ SAS circuit with $n = 16$, this would lead to a total number of 3136 WGX.

Regarding hybrid 2D/3D SAS circuit implementations based on WOP, we again start from the complete bipartite graph $K_{n,n}$ and determine the number of WOP by subtracting the maximum number of edges that can be realized without crossings (the number of edges in the *spanning maximum planar subgraph*) from the total number of edges. The total number of edges in $K_{n,n}$ is $n^2$, and $4n - 4$ edges can be realized without crossings [30]. The number of missing edges hence amounts to

$$\mu_{n,n}^{(\text{surf})} = n^2 - (4n - 4) = (n-2)^2 \qquad (3)$$

and equals the number of WOP necessary to complete the SAS circuit, assuming that each WOP can cross an arbitrary number of planar waveguides, and that crossings of 3D WOP can be avoided, see Appendix B for more details. Note that the length of a WOP is only limited by the write field size of the two-photon lithography system, which currently amounts to approximately 500 µm × 500 µm. In the future, these limitations may be overcome by high-precision stitching of structures that extend across several write fields. Using Eq. (3), we calculate a total number of 196 WOP for an SAS circuit with $n = 16$, which is considerably smaller than the corresponding number of WGX. A comparison of the scaling of WGX and WOP numbers for increasing port count $n$ is given in the second and third column of Table 1.

As a further step of the circuit layout optimization, we may split up the $1 \times n$ and the $n \times 1$ switches at the input and the output into binary trees (BT) of $1 \times 2$ and $2 \times 1$ switches, see Fig. 2(d). This allows to reduce the number of WOP to

$$\mu_{n,n}^{(\text{surf, BT})} = \left\lceil \frac{(n-2)^2}{2} \right\rceil, \qquad (4)$$

see Appendix B for an explanation. In the last relation, $\lceil x \rceil$ denotes the ceiling function. The associated WOP numbers for increasing port counts $n$ are indicated in the fourth column of Table 1. Note that the same technique with BT switches may also be applied to the single-layer SAS circuit architecture as illustrated in Fig. 2(c). For $n = 4$, we could not find a layout that reduces the number of WGX as compared to the implementation with LE switches. Note that the SAS circuit with BT switches is not any more a complete bipartite graph $K_{n,n}$ – a general analysis of this circuit topology is subject to ongoing investigations. Note further that for increasing port counts $n$ of the SAS circuit with LE switches, the number of WGX increases with $n^4/16$, whereas the number of WOP of the SAS circuit with BT switches increases only in proportion to $n^2/2$. As a consequence, the number of WOP in a 16 × 16 SAS circuit with BT switches is nearly two orders of magnitude smaller than the number of WGX with LE switches, and for a 64 × 64 SAS, the numbers differ by nearly four orders of magnitude, see Table 1.

Besides the total number of WGX or WOP in the circuit, the maximum number of such elements along any optical path through the circuit is an important figure of merit. For the single-layer implementation of the SAS circuit with LE switches, the biggest number of WGX along an optical path amounts to

$$\xi_{n,n}^{(\text{surf})} = \left( \left\lceil \frac{n}{2} \right\rceil - 1 \right)^2, \qquad (5)$$

which scales with $n^2/4$ for large $n$, see Appendix B for details. The corresponding numbers for increasing port counts $n$ are given in the fifth column of Table 1. For an SAS circuit with $n = 16$, this leads to up to 49 WGX along a single optical path. In contrast to that, the number of WOP can be kept to at most one along each path, see last column of Table 1.

**Table 1. Quantitative comparison of surface-coupled $n \times n$ switch-and-select (SAS) circuit implementations based on WGX in single-layer circuits and on WOP in hybrid 2D/3D photonic integration.**

| SAS ($n \times n$) | Total number | | | Maximum number along any optical path | |
|---|---|---|---|---|---|
| | WGX (LE) | WOP (LE) | WOP (BT) | WGX (LE) | WOP (LE & BT) |
| $4 \times 4$ | 4 | 4 | 2 | 1 | 1 |
| $8 \times 8$ | 144 | 36 | 18 | 9 | 1 |
| $16 \times 16$ | 3136 | 196 | 98 | 49 | 1 |
| $32 \times 32$ | 57 600 | 900 | 450 | 225 | 1 |
| $64 \times 64$ | 984 064 | 3 844 | 1 922 | 961 | 1 |

The total number of WGX increases approximately in proportion to $n^4/16$, whereas the number of WOP scales with $n^2$ for the case of lumped-element (LE) switches, and with $n^2/2$ in case the switches are decomposed into binary trees (BT) of $1 \times 2$ and $2 \times 1$ switches. The maximum number of WGX along any optical path increases approximately in proportion to $n^2/4$ for the case of LE switches, whereas the maximum number of WOP along any optical path amounts to 1 in both cases of LE and BT switches.

## 4. Device design, fabrication and experimental characterization

To demonstrate the viability of the WOP concept, we realized a $4 \times 4$ SAS device, similar to the one illustrated in Fig. 2(d), featuring two WOP. The device consists of four $1 \times 4$ switches at the input and four $4 \times 1$ switches at the output. Each of the $1 \times 4$ switches is realized as a BT of three $1 \times 2$ switches, and the same technique is applied to the $4 \times 1$ switches. In general, for realizing a $1 \times n$ switch as a BT, we need $(n – 1)$ $1 \times 2$ switches, each of which consists of a Mach-Zehnder interferometer (MZI) comprising two multi-mode interference (MMI) couplers and a pair of thermal phase shifters in the MZI arms. In total, there are $2n(n – 1) = 24$ MZI and $24 \cdot 2 = 48$ phase shifters, leading to 48 signal pads and a common ground for the electrical control signals. Note that activating one of the two phase shifters of each MZI is sufficient for switching – the second phase shifter has only been implemented for better balancing of the MZI arms. We use surface coupling by grating couplers (GC). One of the WOP bridges three, and the other bridges four SiP waveguides spaced by 3.5 µm, see Fig. 2(d) and Fig. 3(c). The footprint of a single WOP amounts to approximately $15 \times 160$ µm$^2$, including two 50 µm-long tapers for coupling the WOP to the SiP waveguides. This is more than an order of magnitude smaller than previously demonstrated overpasses realized by direct laser inscription of low-index contrast 3D-waveguides into glass matrices [31].

For switching, each of the possible input-output connections can be established by activating four phase shifters: Two phase shifters at the BT at the input are used to switch to the targeted output, and another two phase shifters are needed at the BT at the output to select the input. For an $n \times n$ SAS circuit with $n = 4$, accessing the full set of $n! = 24$ switch states would require to operate one phase shifter in each of the 24 MZI. To establish a specific switch state, i.e., a specific set of connections between input and output ports, it is sufficient to simultaneously operate a maximum of $2n \lceil \log_2 n \rceil = 16$ phase shifters, while the remaining phase shifters along unused optical paths are idle. In the experiment, we use a multi-channel current source that we can flexibly connect to the 16 relevant pads out of the overall set of 48 phase shifters. The electrical connection to the chip is established through two multi-contact probe wedges (MCW), see Fig 3(a) and 3(b), each one with 15 DC probes. For each of these wedges, twelve probes connect to the phase shifters, two probes are used for the common ground connection pads on the chip, and one probe is left idle. From the $n^2 = 16$ optical paths connecting the various inputs and outputs of the switch, four paths contain one of the two WOP, see Fig. 2(d).

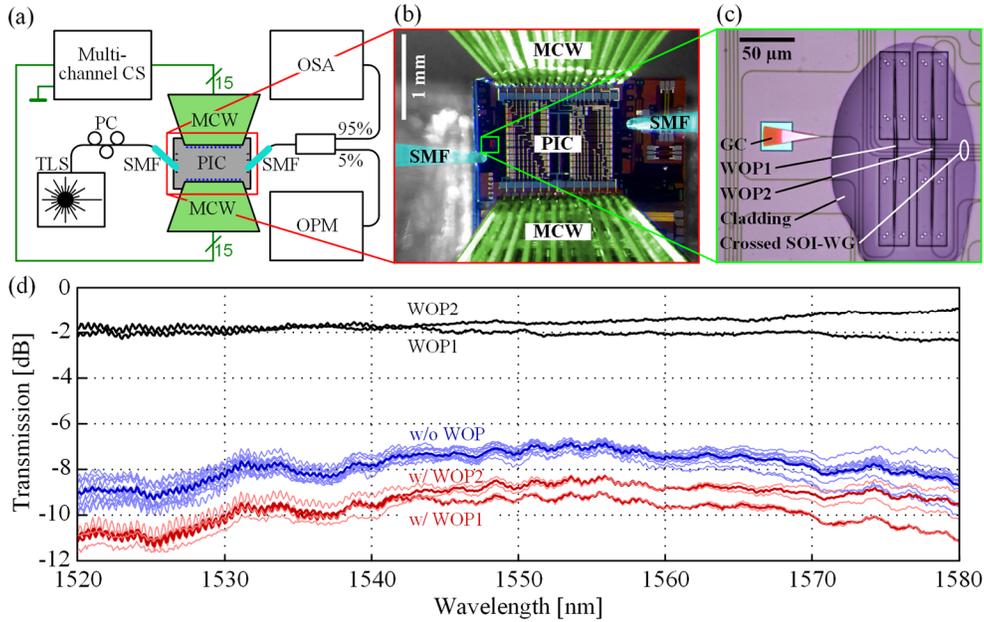

**Fig. 3.** Experimental demonstration of the 4 × 4 SAS with WOP. The layout of the SAS circuit is similar to the one depicted in Fig. 2(d). **(a)** Experimental setup. A multi-channel current source (CS) is used to drive different subsets of 16 out of the overall 24 optical 1 × 2 and 2 × 1 MZI switches via two multi-contact probe wedges (MCW). This allows testing of all 16 possible optical paths that connect the various input and output ports of the 4 × 4 SAS PIC. A tunable laser source (TLS) and a polarization controller (PC) are used to generate continuous-wave (CW) test signals that are launched to the various ports of the SAS PIC via a single-mode fiber (SMF) and grating couplers (GC). Each of the four optical outputs can be probed by another SMF, and the output signal is analyzed with an optical power meter (OPM) and an optical spectrum analyzer (OSA) that allows to perform a wavelength sweep that is synchronized with the TLS. **(b)** Microscope image of the SAS PIC with electrical and optical connections. **(c)** Microscope image of two waveguide overpasses (WOP), which bridge three and four SiP strip waveguides, respectively. A low-index cladding material is locally deposited with high precision to cover the printed WOP without blocking the nearby grating couplers. **(d)** Transmission spectra of various optical paths through the switch. **Pale blue:** Transmission spectra of 12 optical paths through the SAS PIC that do not contain any WOP (w/o WOP). **Bright blue:** Average transmission of the 12 paths w/o WOP. **Pale red:** Transmission spectra of two sets of two optical paths each, each set containing the same WOP (w/ WOP1; w/ WOP2). **Bright red:** Average transmission of each of the two sets w/ WOP. **Black**: Transmission spectra of WOP1 and WOP2.

To characterize the performance of our SAS PIC, we measured the transmission spectra of all 16 optical paths, see Fig. 3(d). To eliminate the fiber-chip coupling losses, we use a reference structure composed of two GC that are connected by a short on-chip waveguide. The GC are not optimized and show maximum transmission at a wavelength of 1560 nm with a fiber-chip coupling loss of approximately 6.3 dB per coupling interface. For each path, we measure the transmission as a function of wavelength, and we correct the data to eliminate the fiber-chip coupling losses. In Fig. 3(d), the transmission spectra of the 12 optical paths without WOP is displayed in pale blue, and the bright blue trace corresponds to the average insertion loss of the 12 paths. At 1550 nm, the average on-chip loss of the paths without WOP amounts to approximately 7 dB and originates from 8 MMI splitters, 4 phase shifters and up to 6.2 mm of on-chip SiP waveguide for each optical path. Using optimized devices on the SiP platform, namely MZI with insertion loss of 0.33 dB [8], waveguides with propagation losses of 0.2 dB/mm, and waveguide lengths of up to only 3 mm, the losses can be reduced to below 2 dB. We also measure the remaining two sets of two paths, each set containing the same WOP – the results are depicted in pale red, and the average for each set is given by a bright red solid line. The insertion losses of the two WOP, indicated as black curves in Fig. 3(d), are extracted from the difference of the bright blue and the two bright red curves by additionally taking into account the different lengths of the on-chip SiP waveguides along the various optical paths. At a wavelength of 1550 nm, we measured insertion losses of 1.6 dB and 1.9 dB for the two WOP. Note that these losses can be reduced to below 1 dB by optimizing the design of the PIC and of the free-form WOP. This was not possible in the current implementation, since the distance between the tips of the tapered on-chip SiP waveguides and the edge of the opening in the $SiO_2$ cladding layer was chosen too small during the design of the PIC. This forces the WOP to have a sharp bend upwards and a sharp bend at its top, and hence leads to noticeable radiation losses.

We also measured the crosstalk from a WOP to one of the SiP waveguides underneath. To this end, we first maximized the optical transmission of two paths through the SAS PIC, where the first path ("Path 1") contains the WOP while the second path ("Path 2") contains one of the SiP waveguides underneath. We then launch a strong CW signal into the input of Path 1, and we connect highly sensitive power detectors to the output of both Path 1 and Path 2. Path 1 and Path 2 are marked in green and in blue, respectively, in Fig. 2(d), and the arrows indicate the direction of light propagation for the crosstalk measurement. To isolate the crosstalk contribution of spurious substrate modes excited at the input grating coupler from the impact of the WOP, we modulated the drive current of MZI right before the WOP ("MZI1", marked in green) with a sinusoidal signal at a distinct lock-in frequency of $f_{LI} = 10$ kHz. We then used a lock-in amplifier to measure the RMS values of the optical power fluctuations at this modulation frequency both at the output of Path 1 and at the output of Path 2. The crosstalk is obtained by calculating the ratio of the two lock-in signals and amounts to $-75$ dB at a wavelength of 1550 nm. Note that this crosstalk figure does not account for differences in on-chip loss between the point where the crosstalk is generated and the output GC of Path 1 and Path 2. Also note that this value very likely represents an upper limit for the WOP crosstalk, since it also contains contributions of other on-chip elements such as waveguide bends and lossy MMI couplers that follow MZI1.

The overall footprint of our current SAS circuit amounts to approximately $1.8 \times 1.4$ mm², mainly dictated by the rather bulky 500 µm-long thermo-optic phase shifters and the associated electric contact pads. This footprint can be reduced by using MZI switch modules based on ultra-compact liquid-crystal phase shifters, which can provide phase shifts in excess of $\pi$ for a length of less than 50 µm [32,33]. Regarding scalability of the WOP to large numbers of crossed waveguides, we have performed simulations of 3D polymer waveguides comparable to WOP in our previous work [25], finding that the insertion loss is dominated by the coupling to the SiP waveguide rather than by the length of the polymer waveguide section. Therefore, assuming an optimized WOP trajectory, increasing the WOP length should not lead to significantly higher losses. Each additionally crossed SiP waveguide increases the WOP length by approximately 3.5 µm, which is dictated by the minimum spacing between the SiP waveguides that is needed to avoid crosstalk between them. Further reduction of the spacing can be achieved by using different SiP waveguide widths to avoid crosstalk [34]. Regarding very complex circuit topologies, the WOP footprint may hence scale very well.

## 5. Summary

We introduced a concept for realizing PIC with non-planar topologies. Planar waveguide crossings (WGX) are replaced by 3D-printed freeform waveguide overpasses (WOP). We demonstrate the viability of the approach using a silicon photonic $4 \times 4$ switch-and-select (SAS) structure. Our theoretical analysis shows that the number of crossings for an $n \times n$ SAS device realized using surface couplers scales with $n^4/16$, while the number of required WOP scales with $n^2/2$. We believe that the results may offer an attractive path towards highly complex PIC with non-planar topologies.

## Appendix A: Graph theory

In this section, we shortly summarize a few *definitions* from graph theory that are used in the graph-theoretical analysis of SAS circuits in Section 3 and Appendices B and C.

(1) A *graph* $G(N,E)$ is defined as an ordered pair consisting of a set of *vertices* $N$ and a set of *edges* $E$, which are two-element subsets of $N$ (one edge connects two vertices). The number of vertices and edges is $|N|$ and $|E|$, respectively. The notation $|X|$ denotes the *cardinality* (number of elements) of a set $X$.

(2) A *bipartite graph* $G(M,N,E)$ consists of two sets of vertices $M$ and $N$ and a set of edges $E$, such that there are no edges between two vertices that are in the same set.

(3) In a *complete graph* $G(N,E)$, each vertex of set $N$ is connected by an edge to all other vertices of the same set. The number of vertices is $|N|=n$, and the number of edges is $|E|=n(n-1)/2$. Such a graph is denoted by $K_n$.

(4) In a *complete bipartite graph* $G(M,N,E)$, each vertex of set $M$ is connected by an edge to each vertex of the second vertex set $N$. The number of vertices is $|M|+|N|=m+n$, and the number of edges is $|E|=mn$. Such a graph is denoted by $K_{m,n}$.

(5) A *planar graph* can be drawn in a plane without *edge crossings*. From Kuratowski's theorem [35], it follows that a complete graph $K_n$ is planar if $n \leq 4$, and a complete bipartite graph $K_{m,n}$ is planar if $m \leq 2$ or $n \leq 2$.

(6) A *maximum planar graph* would become a *non-planar* graph by adding one additional edge.

(7) A *plane embedding* is a drawing of a planar graph in a plane without edge crossings.

(8) A *plane embedding* divides the plane into distinct regions called *faces*. All faces are bounded by edges except for the single *outer face* which extends to infinity. In a *maximum planar graph plane embedding*, each face is defined by three edges. In a *bipartite maximum planar graph plane embedding*, each face is defined by four edges.

(9) The *crossing number* $\mathrm{cr}(G)$ of a graph $G$ counts the minimum number of edge crossings, taking into account all possible drawings of $G$ in a plane. The crossing number of a planar graph is zero.

(10) The *outerplanar crossing number* $\mathrm{cr}^*(G)$ of a graph $G$ counts the minimum number of edge crossings, taking into account all possible drawings of $G$ in a plane, such that all vertices of $G$ lie on a closed boundary curve, and all edges of $G$ are drawn inside the area bounded by the boundary curve.

(11) The *local crossing number* $\mathrm{lcr}(G)$ of a graph $G$ is the minimum of the maximum number of crossings along any edge of $G$, taking into account all possible drawings of $G$ in a plane.

(12) The *local crossing number of a graph drawing* counts the maximum number of edge crossings along any edge for that particular drawing.

(13) A *subgraph* of a graph $G$ is a graph consisting of sets of vertices and edges that are subsets of sets of vertices and edges of $G$.

(14) A *spanning maximum planar subgraph* of a graph $G$ is a *maximum planar subgraph* of $G$ that contains all vertices of $G$.

For more information on general graph theory, please refer to [36]. Crossing number problems are discussed in more detail in [37].

## Appendix B: Graph-theoretical analysis of surface-coupled $m \times n$ SAS circuits

As previously mentioned, a surface-coupled $m \times n$ circuit with $1 \times n$ and $m \times 1$ LE switches at the input and output ports can be represented by a complete bipartite graph $K_{m,n}$. The conjectured crossing number of $K_{m,n}$ is given by [11]

$$\mathrm{cr}_{\mathrm{conj.}}(K_{m,n}) = \eta_{m,n}^{(\mathrm{surf})} = \left\lfloor \frac{m}{2} \right\rfloor \left\lfloor \frac{n}{2} \right\rfloor \left\lfloor \frac{m-1}{2} \right\rfloor \left\lfloor \frac{n-1}{2} \right\rfloor, \tag{B1}$$

which in case $m = n$ reduces to Eq. (2). For a complete bipartite graph $K_{m,n}$, the construction of a drawing that results in the conjectured minimum number of crossings given by Eq. (B1) is proposed in [11] and illustrated in Fig. 4(a) for the case of $K_{5,5}$. In a first step, all vertices of set $M$ are placed on the $x$-axis, whereas the vertices of set $N$ are placed on the $y$-axis of the 2D Cartesian coordinate system. This placement is done such that the number of vertices on both positive and negative parts of the $x$ and $y$-axes is as much equal as possible. Achieving exactly equal numbers is possible only if $m$ and $n$ are even – if any of them is odd, we will put one vertex more on the positive side of the corresponding axis. Therefore, the $x$-coordinates of the vertices of set $M$ are, $-\lfloor m/2 \rfloor, -\lfloor m/2 \rfloor+1, \ldots, -1, 1, \ldots, \lceil m/2 \rceil$, and the corresponding vertices are labelled with $v_{M,-\lfloor m/2 \rfloor}, v_{M,-\lfloor m/2 \rfloor+1}, \ldots, v_{M,-1}, v_{M,1}, \ldots, v_{M,\lceil m/2 \rceil}$. Similarly, the $y$-coordinates of the vertices of set $N$ are, $-\lfloor n/2 \rfloor, -\lfloor n/2 \rfloor+1, \ldots, -1, 1, \ldots, \lceil n/2 \rceil$, and the corresponding vertices are labelled with $v_{N,-\lfloor n/2 \rfloor}, v_{N,-\lfloor n/2 \rfloor+1}, \ldots, v_{N,-1}, v_{N,1}, \ldots, v_{N,\lceil n/2 \rceil}$. Finally, all vertices of set $M$ are connected by $mn$ line segments to all vertices of set $N$.

In order to find the local crossing number of such drawing it is enough to analyze the 1st quadrant of the 2D Cartesian system, since it contains the largest number of vertices and edges, and since all edges are completely drawn in single quadrants. The two edges that cross the largest number of other edges, $\{v_{N,\lceil n/2 \rceil}, v_{M,1}\}$ and $\{v_{N,1}, v_{M,\lceil m/2 \rceil}\}$, are drawn in blue in Fig 4(a). It can be easily seen that edge $\{v_{N,\lceil n/2 \rceil}, v_{M,1}\}$ must cross all edges that connect $\lceil n/2 \rceil-1$ vertices $v_{N,1}, \ldots, v_{N,\lceil n/2 \rceil-1}$ to $\lceil m/2 \rceil-1$ vertices $v_{M,2}, \ldots, v_{M,\lceil m/2 \rceil}$. Similarly, edge $\{v_{N,1}, v_{M,\lceil m/2 \rceil}\}$ must cross all edges that connect $\lceil n/2 \rceil-1$ vertices $v_{N,2}, \ldots, v_{N,\lceil n/2 \rceil}$ to $\lceil m/2 \rceil-1$ vertices $v_{M,1}, \ldots, v_{M,\lceil m/2 \rceil-1}$. Therefore, the local crossing number of this drawing is

$$\mathrm{lcr}_{\mathrm{conj.\,drawing}}(K_{m,n}) = \xi_{m,n}^{(\mathrm{surf})} = \left(\left\lceil \frac{m}{2} \right\rceil - 1\right)\left(\left\lceil \frac{n}{2} \right\rceil - 1\right). \tag{B2}$$

For $m = n$, this reduces to Eq. (5).

To analyze the number of necessary WOP, we introduce a term *3D edge*, which is an edge that is not restricted to the plane but can be routed in 3D, and we will use it to model a WOP. A WOP does not directly connect two optical devices on the PIC, but rather links two ends of two planar waveguides, each of which is connected to an optical device at its respective other end. The connections of WOP and planar waveguides are an analog to metallic vias that connect metallic wires in different layers of an electric printed circuits board (PCB). In the graph representation, a WOP is modelled by a 3D edge that does not directly connect to two vertices on the plane, but links two planar edges, each of which is connected to another vertex at its respective other end. In order to estimate the number of necessary 3D edges, we first construct a spanning maximum planar subgraph of $K_{m,n}$, which has $2m + 2n - 4$ edges [30]. We do it by connecting each of the vertices $v_{M,-\lfloor m/2 \rfloor}, v_{M,\lceil m/2 \rceil}, v_{N,-1}$, and $v_{N,1}$ to each vertex of the opposite set, see Fig. 4(b). The remaining

$$\mu_{m,n}^{(\mathrm{surf})} = mn - (2m + 2n - 4) = (m-2)(n-2) \tag{B3}$$

edges can be realized using 3D edges. For $m = n$, Eq. (B3) reduces to Eq. (3). The concept is illustrated in Fig. 4(b) for the case of $K_{5,5}$. The edges of the spanning maximum planar subgraph are depicted in blue, the 3D edges in red (dashed), while the planar edges that connect the 3D edges to the vertices are depicted in black. The red dashed lines are, in fact, vertical projections of 3D edges on the 2D drawing plane.

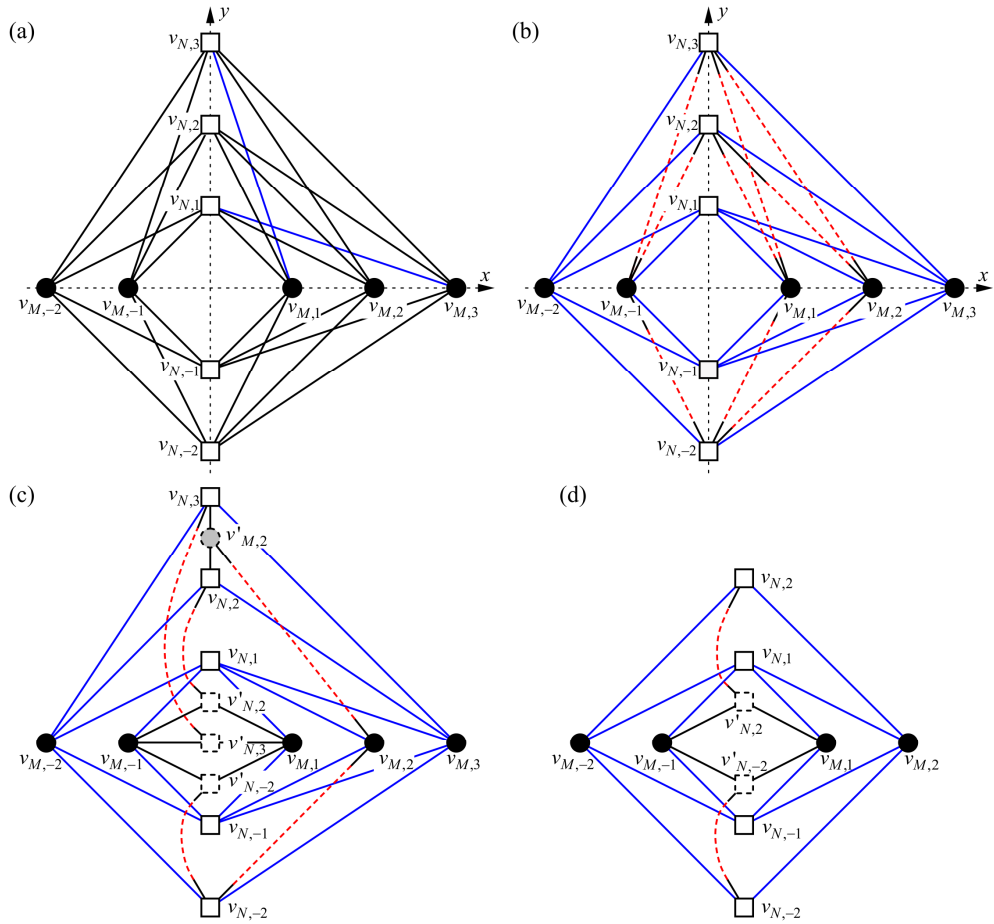

**Fig. 4.** Different graph drawings of a surface-coupled 5 × 5 and 4 × 4 SAS circuit: **(a)** Graph drawing of a 5 × 5 SAS circuit where 1 × 5 and 5 × 1 switches at the input and output ports are realized as LE. The circuit is modeled by a complete bipartite graph $K_{5,5}$, and the arrangement of vertices of sets $M$ and $N$ is such that the drawing results in the number of crossings equal to the conjectured crossing number given by Eq. (B1). The two edges depicted in blue are the edges with the maximum number of crossings, which determine the local crossing number of this particular graph drawing, as given by Eq. (B2). **(b)** Planar-edge-crossing-free graph representation of the same circuit. The edges depicted in blue represent a spanning maximum planar subgraph of $K_{5,5}$. The remaining edges are realized with help of 3D edges (representing WOP) depicted as dashed red lines, which are routed outside the plane of the drawing and avoid crossings with the planar edges. Each 3D edge connects to a pair of planar edges depicted in black, that are linked to vertices at the respective other end. **(c)** If the 1 × 5 and 5 × 1 switches at the input and output ports are realized as BT of 1 × 2 and 2 × 1 switches, the number of necessary 3D edges can be reduced by splitting the vertices of the original $K_{5,5}$ and placing them into appropriate faces of the spanning maximum planar subgraph depicted in blue. The white dashed squares represent 1 × 2 switches, while the dashed circle with gray filling represents a 2 × 1 switch. This approach allows to replace a pair of 3D edges by a single one. In case where both $m$ and $n$ are odd, the number of missing edges is also odd, and one missing edge (in this case $\{v_{N,-2}, v_{M,2}\}$) must be realized with help of one single 3D edge. **(d)** Graph drawing of a 4 × 4 SAS circuit, analogous to the case described in (c). In case at least one of the numbers $m$ or $n$ is even, the number of 3D edges can be reduced by a factor of 2 compared to the case when 1 × $n$ and $m$ × 1 switches are realized as LE. For our experimental demonstration, we used the PIC layout displayed in Fig. 2(d), which was obtained in a similar way as Fig. 4(d), with the difference that the auxiliary vertices (1 × 2 switches) in Fig. 2(d) were placed in the outer face of the spanning maximum planar subgraph rather than in its inner face, as displayed here.

Note that the planar projections of the 3D edges on the drawing plane may cross each other. This, however, does not mean that the two 3D edges cross in 3D space – two freeform WOP can always be 3D-printed such that one passes over the other, and the corresponding 3D edges can be routed analogously. Furthermore, by appropriate routing of the planar and 3D edges, the crossings of projections of 3D edges on the drawing plane can be avoided. Fig. 4(b) shows how a possible crossing of projections of two 3D edges between pairs of vertices $\{v_{N,2}, v_{M,2}\}$ and $\{v_{N,3}, v_{M,1}\}$ has been avoided by making the planar waveguide that connects $v_{N,2}$ to the corresponding 3D edge sufficiently long such that it passes underneath the 3D edge between the pair of vertices $\{v_{N,3}, v_{M,1}\}$. We believe that this approach might be generalized to avoid crossings of projections of 3D edges for general complete bipartite graphs $K_{m,n}$ – a general proof would need further investigation and is beyond the scope of this paper.

If the $1 \times n$ and $m \times 1$ switches at the input and output ports are realized as BT of $1 \times 2$ and $2 \times 1$ switches rather than as LE, we can further reduce the number of 3D edges. We will split the analysis into two cases: when both $m$ and $n$ are odd, and when at least one of them is even. Furthermore, we will only analyze cases where both $m$ and $n$ are $\geq 3$ since otherwise, according to Kuratowski's theorem [35], the complete bipartite graph $K_{m,n}$ is planar. If both $m$ and $n$ are odd, we do the following steps:

**Step 1:** Construct the spanning maximum planar subgraph of $K_{m,n}$ as described above. The edges of this subgraph are depicted in blue in Fig. 4(c) for the case of $K_{5,5}$ ($m = n = 5$). This subgraph has all faces determined by four vertices (two from set $M$ and two from set $N$) and four edges. There are ($m - 3$) vertices of set $M$ whose x-coordinates lie between $-\lfloor m/2 \rfloor + 1$ and $\lceil m/2 \rceil - 2$ inclusive, and they can be divided into ($m - 3$)/2 distinct two-element subsets of adjacent vertices (because $m - 3$ is even, and therefore divisible by two): $\{v_{M,-\lfloor m/2 \rfloor+1}, v_{M,-\lfloor m/2 \rfloor+2}\}, \{v_{M,-\lfloor m/2 \rfloor+3}, v_{M,-\lfloor m/2 \rfloor+4}\}, \ldots, \{v_{M,\lceil m/2 \rceil-3}, v_{M,\lceil m/2 \rceil-2}\}$. Each of these ($m - 3$)/2 pairs of vertices of set $M$ together with the pair of vertices $\{v_{N,-1}, v_{N,1}\}$ of set $N$, define ($m - 3$)/2 faces: $\{v_{M,-\lfloor m/2 \rfloor+1}, v_{N,-1}, v_{M,-\lfloor m/2 \rfloor+2}, v_{N,1}\}, \{v_{M,-\lfloor m/2 \rfloor+3}, v_{N,-1}, v_{M,-\lfloor m/2 \rfloor+4}, v_{N,1}\}, \ldots, \{v_{M,\lceil m/2 \rceil-3}, v_{N,-1}, v_{M,\lceil m/2 \rceil-2}, v_{N,1}\}$. For $m = 3$ there are no such faces. For $m = 5$, there is only one such face $\{v_{M,-\lfloor m/2 \rfloor+1}, v_{N,-1}, v_{M,\lceil m/2 \rceil-2}, v_{N,1}\} = \{v_{M,-1}, v_{N,-1}, v_{M,1}, v_{N,1}\}$, see Fig. 4(c). Note that the results of the expressions $-\lfloor m/2 \rfloor + i$ and $\lceil m/2 \rceil - j$ in the subscripts of labels of vertices of set $M$ indicate the x-coordinates of the vertices. Since there is no vertex at $x = 0$, not a single expression is allowed to result in zero. Therefore, we restrict the values of integers $i$ and $j$ to $i = 1, 2, \ldots, \lfloor m/2 \rfloor - 1$, and $j = \lceil m/2 \rceil - 1, \lceil m/2 \rceil - 2, \ldots, 2$ (the expression $-\lfloor m/2 \rfloor + i$ is used for vertices on the negative side of the x-axis, while the expression $\lceil m/2 \rceil - j$ is used for vertices on the positive side of the x-axis).

**Step 2:** Let us put an auxiliary vertex $v'_{N,\lceil n/2 \rceil}$ inside the face defined by vertices $\{v_{M,-\lfloor m/2 \rfloor+1}, v_{N,-1}, v_{M,-\lfloor m/2 \rfloor+2}, v_{N,1}\}$. We can connect the auxiliary vertex $v'_{N,\lceil n/2 \rceil}$ to vertex $v_{N,\lceil n/2 \rceil}$ with a 3D edge, and the same auxiliary vertex to vertices $v_{M,-\lfloor m/2 \rfloor+1}$ and $v_{M,-\lfloor m/2 \rfloor+2}$ with two planar edges. The auxiliary vertex is the place where we put a $2 \times 1$ switch which is a part of the BT $m \times 1$ switch at the vertex $v_{N,\lceil n/2 \rceil}$. In this way, we can replace two 3D edges that would otherwise separately connect vertex $v_{N,\lceil n/2 \rceil}$ to vertices $v_{M,-\lfloor m/2 \rfloor+1}$ and $v_{M,-\lfloor m/2 \rfloor+2}$. The auxiliary vertex $v'_{N,\lceil n/2 \rceil}$ and the two planar edges that connect it to vertices $v_{M,-\lfloor m/2 \rfloor+1}$ and $v_{M,-\lfloor m/2 \rfloor+2}$ splits the original face $\{v_{M,-\lfloor m/2 \rfloor+1}, v_{N,-1}, v_{M,-\lfloor m/2 \rfloor+2}, v_{N,1}\}$ into two faces $\{v_{M,-\lfloor m/2 \rfloor+1}, v_{N,-1}, v_{M,-\lfloor m/2 \rfloor+2}, v'_{N,\lceil n/2 \rceil}\}$ and $\{v_{M,-\lfloor m/2 \rfloor+1}, v_{N,1}, v_{M,-\lfloor m/2 \rfloor+2}, v'_{N,\lceil n/2 \rceil}\}$. We put an additional auxiliary vertex $v'_{N,\lceil n/2 \rceil-1}$ to any of the two new faces, and we connect it to vertex $v_{N,\lceil n/2 \rceil-1}$ with a 3D edge and to vertices $v_{M,-\lfloor m/2 \rfloor+1}$ and $v_{M,-\lfloor m/2 \rfloor+2}$ with two planar edges. We repeat the procedure for all vertices of set $N$, except for vertices $v_{N,-1}$ and $v_{N,1}$, which are already connected to all vertices of set $M$. In this way, we connect both vertices of the pair $\{v_{M,-\lfloor m/2 \rfloor+1}, v_{M,-\lfloor m/2 \rfloor+2}\}$ to all vertices of set $N$. We apply the same algorithm to connect the pairs of vertices $\{v_{M,-\lfloor m/2 \rfloor+3}, v_{M,-\lfloor m/2 \rfloor+4}\}, \ldots, \{v_{M,\lceil m/2 \rceil-3}, v_{M,\lceil m/2 \rceil-2}\}$ to all vertices of set $N$. This step has been illustrated in Fig. 4(c) where auxiliary vertices $v'_{N,3}$, $v'_{N,2}$, and $v'_{N,-2}$ have been placed inside the face $\{v_{M,-1}, v_{N,-1}, v_{M,1}, v_{N,1}\}$, connected to vertices $v_{N,3}$, $v_{N,2}$, and $v_{N,-2}$ by 3D edges, respectively, and to $v_{M,-1}$ and $v_{M,1}$ by planar edges. For $m = 3$, Step 2 is skipped.

**Step 3:** To this point, we connected all vertices of set $M$ to all vertices of set $N$, except for vertex $v_{M,\lceil m/2 \rceil-1}$ that is connected only to $v_{N,-1}$ and $v_{N,1}$ and still needs to be connected to the remaining $(n - 2)$ vertices of set $N$. There are $\lceil n/2 \rceil - 1$ such vertices on the positive side of the $y$-axis: $v_{N,\lceil n/2 \rceil}, v_{N,\lceil n/2 \rceil-1}, \ldots, v_{N,2}$, and $\lfloor n/2 \rfloor - 1$ on the negative side of the $y$-axis: $v_{N,-2}, v_{N,-3}, \ldots, v_{N,-\lfloor n/2 \rfloor}$. Depending on $n$, one of these two numbers is even, and the other is odd. If $\lceil n/2 \rceil - 1$ is even and $\lfloor n/2 \rfloor - 1$ is odd, then each of the following pairs of vertices $\{v_{N,\lceil n/2 \rceil}, v_{N,\lceil n/2 \rceil-1}\}, \ldots, \{v_{N,3}, v_{N,2}\}, \{v_{N,-2}, v_{N,-3}\}, \ldots, \{v_{N,-\lfloor n/2 \rfloor+2}, v_{N,-\lfloor n/2 \rfloor+1}\}$ together with the pair of vertices $\{v_{M,-\lfloor m/2 \rfloor}, v_{M,\lfloor m/2 \rfloor}\}$ define one face. In each of these faces, we can place one auxiliary vertex $v'_{M,\lceil m/2 \rceil-1}, v''_{M,\lceil m/2 \rceil-1}, v'''_{M,\lceil m/2 \rceil-1}, \ldots$, see Fig. 4(c), where there is only one such auxiliary vertex $v'_{M,\lceil m/2 \rceil-1} = v'_{M,2}$. Each of these auxiliary vertices can be connected to $v_{M,\lceil m/2 \rceil-1}$ by a 3D edge and to the respective pair of vertices of set $N$ (that define the face in which the auxiliary vertex is placed) by two planar edges. After this step, there will be only one missing edge between vertices $v_{M,\lceil m/2 \rceil-1}$ and $v_{N,-\lfloor n/2 \rfloor}$, and we directly connect these two vertices by a single 3D edge, see Fig. 4(c). Similarly, if $\lceil n/2 \rceil - 1$ is odd and $\lfloor n/2 \rfloor - 1$ is even, we can group the vertices of set $N$ into pairs as $\{v_{N,\lceil n/2 \rceil-1}, v_{N,\lceil n/2 \rceil-2}\}, \ldots, \{v_{N,3}, v_{N,2}\}, \{v_{N,-2}, v_{N,-3}\}, \ldots, \{v_{N,-\lfloor n/2 \rfloor+1}, v_{N,-\lfloor n/2 \rfloor}\}$ which would define faces together with the pair of vertices $\{v_{M,-\lfloor m/2 \rfloor}, v_{M,\lfloor m/2 \rfloor}\}$. After placing and connecting auxiliary vertices as described, the only missing edge would be between $v_{M,\lceil m/2 \rceil-1}$ and $v_{N,\lceil n/2 \rceil}$, and we would connect them by one 3D edge.

The case when at least one of the numbers $m$ and $n$ is even is simpler. We can assume without loss of generality that $m$ is even and $n$ is odd. By constructing the spanning maximum planar subgraph of $K_{m,n}$ as described above, we will get a subgraph where each of the following $(m-2)/2$ pairs of vertices $\{v_{M,-\lfloor m/2 \rfloor+1}, v_{M,-\lfloor m/2 \rfloor+2}\}, \{v_{M,-\lfloor m/2 \rfloor+3}, v_{M,-\lfloor m/2 \rfloor+4}\}, \ldots, \{v_{M,\lceil m/2 \rceil-2}, v_{M,\lceil m/2 \rceil-1}\}$ together with the pair of vertices $\{v_{N,-1}, v_{N,1}\}$ define one face. After performing Step 2 as described above, we will connect all vertices of set $M$ to all vertices of set $N$. Fig. 4(d) shows an example of the result of the algorithm for the case of $K_{4,4}$.

The described algorithm allows to replace two missing planar edges by one 3D edge. The number of necessary 3D edges hence amounts to

$$\mu_{m,n}^{(\text{surf, BT})} = \left\lceil \frac{(m-2)(n-2)}{2} \right\rceil, \tag{B4}$$

which reduces to Eq. (4) for $m = n$. The ceiling function in Eq. (B4) is used to include the case when the number of missing edges is odd (both $m$ and $n$ are odd) and not divisible by two (one single missing edge needs to be realized with one single 3D edge). This algorithm is just an example and not the unique way of constructing a layout that results in the number of 3D edges given by Eq. (B4): For example, in Step 1 we could construct the spanning maximum planar subgraph in a different way and then modify Steps 2 and 3 accordingly.

It should be pointed out that Eq. (B4) does not necessarily give the minimum number of necessary 3D edges, but an upper bound. In our construction we started from the spanning maximum planar subgraph, and we split some vertices in two by introducing auxiliary vertices. We did, however, not show that the spanning maximum planar subgraph of $K_{m,n}$ is the optimal way to start with. We could have also started with a non-maximum planar subgraph and could have used larger split ratio switches $1 \times n'$, $n' < n$ and $1 \times m'$, $m' < m$, and place them in the auxiliary vertices. Furthermore, the graph model of the device where $1 \times n$ and $m \times 1$ switches at the input and output ports are realized as BT of $1 \times 2$ and $2 \times 1$ switches, is not a complete bipartite graph $K_{m,n}$. The crossing number of the SAS circuit realized with such an approach is subject to ongoing investigations.

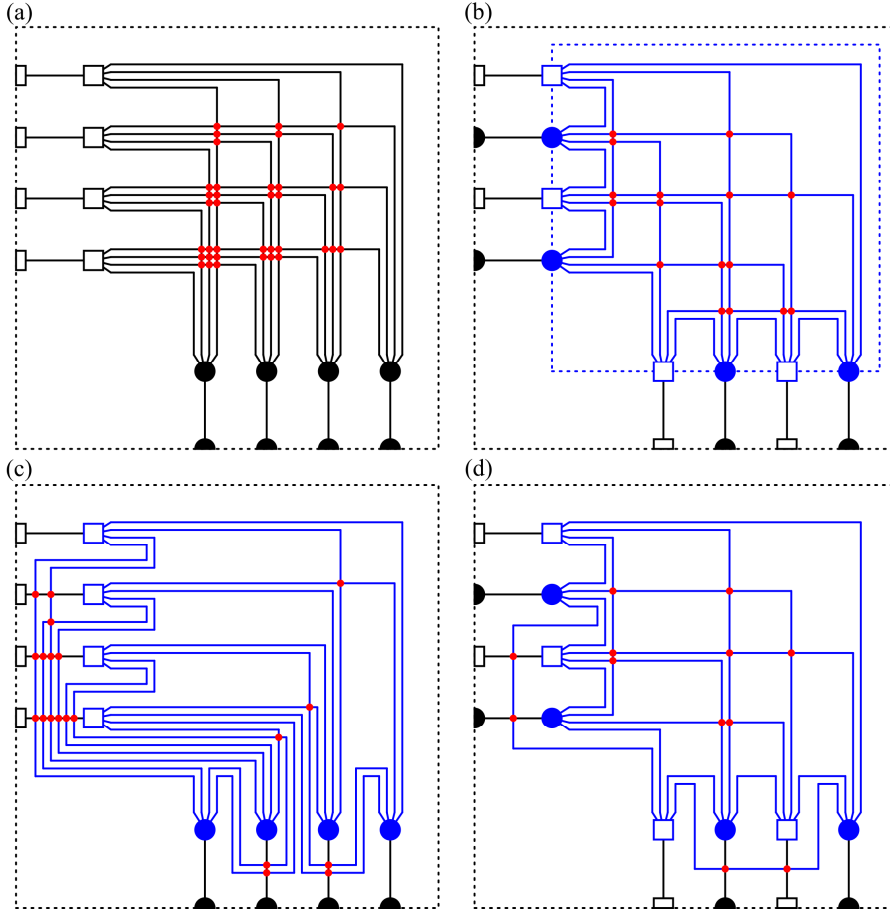

**Fig. 5.** Different graph drawings of a facet-coupled 4 × 4 SAS circuit: A first set of vertices (half circles and rectangles) is used to represent facet-coupled optical input and output ports, and a second kind of vertices (full circles and squares) represents the $1 \times n$ or $m \times 1$ switches. Each port vertex is connected to the associated switch vertex by a graph edge that represents the *access* waveguide **(a)** Simplistic non-optimum graph representation based on the same approach as the surface-coupled SAS circuit shown in Fig. 2(a). Input and output ports are clustered into two groups of neighboring vertices along the chip boundary. For a 4 × 4 SAS circuit, 36 WGX are required. **(b)** By interleaving the input and output ports along the chip boundary, the number of crossings can be reduced, leading to a total number of 16 WGX for a 4 × 4 SAS circuit. **(c)** The number of crossings can also be reduced by allowing routing of waveguides between the ports and the corresponding $1 \times n$ and $n \times 1$ switches, leading to a total number of 20 WGX for the depicted graph drawing. **(d)** Circuit layout obtained by combining interleaving of input and output ports with routing of waveguides between the ports and the corresponding switches, leading to a total number of 12 WGX.

## Appendix C: Facet-coupled SAS circuits

### C.1 Facet-coupled SAS realized in single-layer and hybrid 2D/3D photonic integration

For facet-coupled SAS circuits, all input and output ports are implemented by waveguide facets arranged along the chip boundaries, making it impossible to route waveguides "behind" the ports, i.e., between the ports and the chip boundary. In the graph drawing of the circuit, all vertices representing input and output ports must hence be placed on a closed curve that represents the boundary of the chip surface, and no waveguide (graph edge) routing outside the area enclosed by the curve is allowed. In addition, in contrast to surface coupling, the graph of a facet-coupled SAS is not anymore a complete bipartite graph: For the case of surface coupling, a port and the associated $1 \times n$ or $m \times 1$ switch can be combined into a single vertex, whereas facet-coupled circuits must be represented by a first kind of vertices for the switches and a second kind of vertices for the ports, which must be placed onto the boundary curve. Every port vertex must be connected to the associated switch by a graph edge that represents the *access* waveguide. This results in a *3-partite* graph, which comprises three parties of vertices represented by the ports, the $1 \times n$ switches, and the $m \times 1$ switches, and which is not complete. For a general description of facet-coupled SAS circuits, we can hence not resort any more to the existing theory of complete bipartite graphs. This renders theoretical assessment of the topologies more difficult and requires simplifying assumptions for quantifying the numbers of WGX or WOP. Nevertheless, we believe that non-planar facet-coupled SAS circuits can also greatly benefit from replacing WGX by WOP.

To support this claim, we first consider the basic non-optimized representation of a $4 \times 4$ SAS circuit, see Fig. 5(a). This representation relies on the same simplistic approach as the surface-coupled SAS circuit that is sketched in Fig 2(a). In this approach, each pair of vertices of set $M$ is connected by four edges to each pair of vertices of set $N$, and the four edges make exactly one crossing. The number of crossings is therefore equal to the product of numbers of ways to choose two-element subsets of $M$ and $N$ and amounts to

$$\eta_{m,n}^{(\text{facet, basic})} = \binom{m}{2}\binom{n}{2} = \left(\frac{m(m-1)}{2}\right)\left(\frac{n(n-1)}{2}\right), \tag{C1}$$

which reduces to Eq. (1) for the case of $m = n$. Interleaving the input and output ports along the boundary line allows to reduce this number, see Fig. 5(b). In this case, we can simplify the theoretical assessment to finding the outerplanar crossing number of a complete bipartite graph. This can be seen if we look at the blue dashed line in Fig. 5(b): All vertices representing $1 \times n$ and $m \times 1$ switches are placed on it, and all edges are routed inside the area bounded by it. Note that this implementation is not yet optimal since it does not exploit the possibility to reduce the number of WGX by routing waveguides between the ports and the corresponding $1 \times n$ or $m \times 1$ switches. For the case of $n$ being an integer multiple of $m$, the outerplanar crossing number of a complete bipartite graph $K_{m,n}$ is obtained when the vertices of set $M$ are evenly interleaved between the vertices of set $N$ and amounts to [38]

$$\eta_{m,n}^{(\text{facet})} = \frac{1}{12}n(m-1)(2mn-3m-n). \tag{C2}$$

For the case $m = n$, this reduces to

$$\eta_{n,n}^{(\text{facet})} = \frac{1}{6}n^2(n-1)(n-2), \tag{C3}$$

Which scales with $n^4/6$ for large $n$. The associated numbers of WGX for switches implemented as LE are listed in the second column of Table 2. Further layout optimization steps may involve routing of waveguides between the ports and the corresponding switches, possibly in combination with interleaving of the ports along the chip boundary, see Fig 5(c) and 5(d). Even though we are not aware of any relations specifying the exact crossing numbers of these graphs, we may still use the number of WGX in the associated surface-coupled SAS as a lower bound. This can be understood by observing that both implementations in Fig. 5(c) and 5(d) contain a maximum bipartite subgraph (indicated in blue) which is equivalent to that of the corresponding surface-coupled circuit, Fig. 2(b), and which is complemented by additional crossings caused by the access waveguides. The number of WGX still scales with at least $n^4/16$, see Eq. (2). Similarly to the case of surface-coupled SAS, disaggregating the $1 \times n$ and $m \times 1$ switches into BT of $1 \times 2$ switches might reduce the number of WGX – this aspect is still under investigation. For the remainder of this section, we rely on Eq. (C3) for determining the number of WGX in the facet-coupled $n \times n$ SAS circuit.

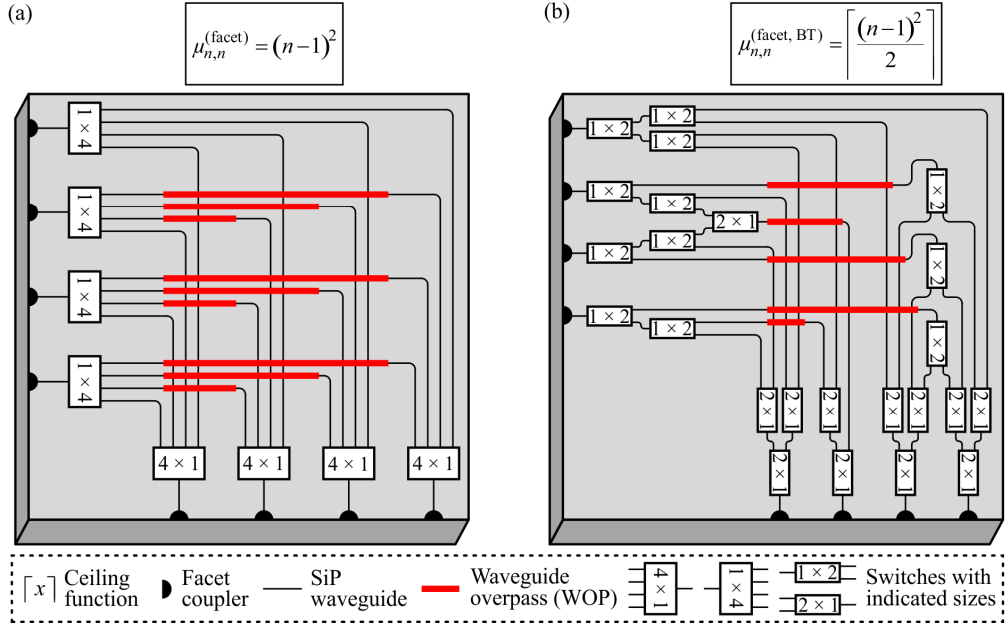

**Fig. 6.** Circuit layouts for facet-coupled 2D/3D hybrid 4 × 4 SAS. **(a)** Simple, but not optimal layout, where the 1 × 4 and 4 × 1 switches have been realized as LE. The relation for $\mu_{n,n}^{(\text{facet})}$ represents the exact number of WOP in this simplistic implementation. **(b)** Best found, but not necessarily optimal layout for the case in which the 1 × 4 and 4 × 1 switches have been realized as BT of 1 × 2 and 2 × 1 switches. The relation for $\mu_{n,n}^{(\text{facet, BT})}$ is an upper bound for the minimum number of WOP.

For 2D/3D hybrid implementations, the number of WOP in facet-coupled SAS circuits was analyzed based on the simplistic layout shown in Fig. 5(a), both for the case of LE switches and for the case of BT cascaded 1 × 2 switches, see Fig. 6(a) and (b). Mathematical details can be found in Appendix C.2. For LE switches, this leads to a total WOP number of

$$\mu_{m,n}^{(\text{facet})} = (m-1)(n-1), \tag{C4}$$

which reduces to $(n-1)^2$ for $m = n$. For BT switches, the number of WOP amounts to

$$\mu_{m,n}^{(\text{facet, BT})} = \left\lceil \frac{(m-1)(n-1)}{2} \right\rceil, \tag{C5}$$

i.e., $\left\lceil (n-1)^2/2 \right\rceil$ for $m = n$. Hence, in both cases, the number of WOP in the facet-coupled hybrid 2D/3D implementation scales much more favorably than the number of WGX in the corresponding single-layer SAS circuit, see third and fourth column of Table 2. Note that this number represents an upper bound for the number of WOP, which might be further reduced by interleaving of ports and by rerouting of connections across the access waveguides, similarly to the case of the surface-coupled planar circuits shown in Fig 2(b), (c), and (d). As in the case of surface coupling, the number of WOP along any optical path through the facet-coupled hybrid 2D/3D circuit is at most 1, whereas the maximum number of WGX along an optical path in a single-layer implementation shown in Fig 5(b) increases in proportion to $n^2/2$. The exact result for the maximum number of WGX along any optical path for this implementation is

$$\xi_{m,n}^{(\text{facet})} = 2 \left\lfloor \frac{n-1}{2} \right\rfloor \left\lceil \frac{n-1}{2} \right\rceil, \tag{C6}$$

see Section C2. The corresponding numbers for $n$ = 4, 8, 16, 32, and 64 are indicated in the fifth and sixth column of Table 2. For the implementations shown in Fig. 5(c) and (d), we cannot provide a formula that describes the minimum number of WGX along an optical path.

**Table 2. Quantitative comparison of $n \times n$ facet-coupled switch-and-select (SAS) circuit implementations based on WGX in single-layer circuits and on WOP in hybrid 2D/3D photonic integration.**

| SAS ($n \times n$) | Total number | | | Maximum number along an optical path | |
|---|---|---|---|---|---|
| | WGX (LE) | WOP (LE) | WOP (BT) | WGX (LE) | WOP (LE & BT) |
| $4 \times 4$ | 16 | 9 | 5 | 4 | 1 |
| $8 \times 8$ | 448 | 49 | 25 | 24 | 1 |
| $16 \times 16$ | 8 960 | 225 | 113 | 112 | 1 |
| $32 \times 32$ | 158 720 | 961 | 481 | 480 | 1 |
| $64 \times 64$ | 2 666 496 | 3 969 | 1 985 | 1984 | 1 |

The total number of WGX (second column) increases approximately in proportion to $n^4/6$, whereas the number of WOP scales with $n^2$ for the case of LE switches (third column) and with $n^2/2$ in case the switches are decomposed into BT of $1 \times 2$ and $2 \times 1$ switches (fourth column). The maximum number of WGX along an optical path increases approximately in proportion to $n^2/2$ for the case of LE switches (fifth column), whereas the maximum number of WOP along an optical path is one in both cases of LE and BT switches (sixth column).

## C.2 Graph-theoretical models and analysis of facet-coupled $m \times n$ SAS circuits

Let us first explain Eq. (C6) which is obtained in case of a drawing of the complete bipartite subgraph $K_{n,n}$ that results in the outerplanar crossing number – the vertices belonging to different vertex sets $M$ and $N$ ($m = n$) are interleaved along the boundary curve. This subgraph is depicted in blue in Fig. 5(b) for $n = 5$. The concept of this layout is illustrated in Fig. 7, which shows graph drawings of a complete bipartite graph $K_{n,n}$ with interleaved vertices of two different vertex sets along the boundary (dashed circular line). Each edge divides the bounded area in two parts, and the largest number of crossings will be on an edge $\{v_{M,i}, v_{N,i}\}$ that divides the area such that the numbers of vertices in both parts are as much equal as possible. If $n$ is odd, it is possible to find an edge that divides the bounded area such that both parts have exactly the same number of vertices; on the other hand, if $n$ is even, one part will have one more vertex of each vertex set than the other part. Both cases are illustrated in Fig. 7 – the edge $\{v_{M,i}, v_{N,i}\}$ is depicted in blue. For both cases, edge $\{v_{M,i}, v_{N,i}\}$ divides the area such that one part contains $\lfloor (n-1)/2 \rfloor$ and the other $\lceil (n-1)/2 \rceil$ vertices of each vertex set. That means that edge $\{v_{M,i}, v_{N,i}\}$ is crossed by $\lfloor (n-1)/2 \rfloor \cdot \lceil (n-1)/2 \rceil$ edges connecting $\lfloor (n-1)/2 \rfloor$ vertices of set $M$ in the first part to $\lceil (n-1)/2 \rceil$ vertices of set $N$ in the second part, and the same number of edges connecting $\lfloor (n-1)/2 \rfloor$ vertices of set $N$ in the first part to $\lceil (n-1)/2 \rceil$ vertices of set $M$ in the second part. From here follows the result of Eq. (C6).

In order to estimate the number of necessary WOP (3D edges), we will use the simplistic layout shown in Fig. 5(a). It is sufficient to consider a drawing of $K_{m,n}$ with all vertices placed on a closed boundary curve, since access waveguides do not have any crossings. We construct a corresponding graph drawing by placing all $m$ vertices of set $M$: $v_{M,1}, v_{M,2}, \ldots, v_{M,m}$ on the $x$-axis of the 2D Cartesian coordinate system in points $x = 1, 2, \ldots, m$, see Fig. 8(a) for an illustration of the case of $K_{4,4}$. Similarly, we place all $n$ vertices of set $N$: $v_{N,1}, v_{N,2}, \ldots, v_{N,n}$ on the $y$-axis in points $y = 1, 2, \ldots, n$. Finally, we connect all vertices of set $M$ to all vertices of set $N$ by $mn$ line segments. The boundary curve can be for example a rectangle that is oriented along the $x$ and the $y$ axis, as depicted in green in Fig. 8(a). The total number of crossings is equivalent to the one given by Eq. (C1) – it takes four edges and one crossing to connect each possible pair of vertices of set $M$ to each possible pair of vertices of set $N$. In order to estimate the number of necessary 3D edges, we first construct a spanning planar subgraph of $K_{m,n}$, by connecting vertices $v_{M,1}$ and $v_{N,n}$ to all vertices of the opposite vertex sets, Fig. 8(b). This subgraph evidently has $m + n - 1$ edges, and the number of missing edges is therefore equal to $mn - (m + n - 1) = (m - 1)(n - 1)$, which leads to Eq. (C4). These edges can be realized with help of 3D edges, illustrated by dashed red lines in Fig. 8(b).

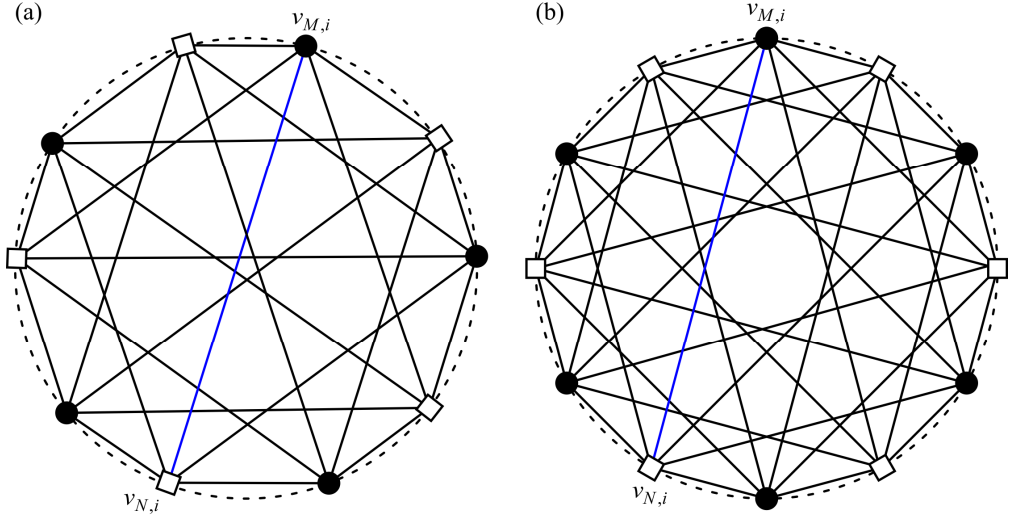

**Fig. 7.** Drawings of complete bipartite graphs $K_{n,n}$ with all vertices placed on the closed boundary curve (dashed circular line), and the vertices of two different sets being interleaved along the boundary. Eq. (C6) gives the local crossing number of such drawing, which occurs along the blue edges that divide the boundary area in two parts such that the number of vertices in both parts is as much balanced as possible. **(a)** In case $n$ is odd (here: $n = 5$), both parts contain the same number of vertices. **(b)** In case $n$ is even (here: $n = 6$), there is one more vertex of each vertex set in one part.

Similarly to the case of surface coupled SAS described in Appendix B, if the $1 \times n$ and $m \times 1$ switches at the input and output ports are BT of $1 \times 2$ and $2 \times 1$ switches, the number of necessary 3D edges reduces. We will split the analysis into two cases: when both $m$ and $n$ are even, and when at least one of them is odd. If both are even, the analysis comprises the following steps:

**Step 1:** Construct a spanning planar subgraph of $K_{m,n}$ as described above. Each pair of vertices $\{v_{M,2}, v_{M,3}\}, \{v_{M,4}, v_{M,5}\}, \ldots, \{v_{M,m-2}, v_{M,m-1}\}$, together with vertex $v_{N,n}$ defines one area, which is bounded by two edges between $v_{N,n}$ and the two vertices of set $M$ and a portion of the $x$-axis between the two vertices of set $M$. This is illustrated on an example of $K_{4,4}$ in Fig. 8(c).

**Step 2:** Put an auxiliary vertex $v'_{N,1}$ inside the area defined by vertices $\{v_{M,2}, v_{M,3}, v_{N,n}\}$, see Fig. 8(c). We can connect the auxiliary vertex $v'_{N,1}$ and vertex $v_{N,1}$ by a 3D edge, and the same auxiliary vertex and vertices $v_{M,2}$ and $v_{M,3}$ by planar edges. Similarly to the case of surface-coupled SAS, we continue adding auxiliary vertices $v'_{N,2}, v'_{N,3}, \ldots, v'_{N,n-1}$, to the same area until we connect all vertices of set $N$ to $v_{M,2}$ and $v_{M,3}$. We then continue with the same procedure for the following areas defined by groups of three vertices: $\{v_{M,4}, v_{M,5}, v_{N,n}\}, \ldots, \{v_{M,n-2}, v_{M,n-1}, v_{N,n}\}$.

**Step 3:** In this fashion, we will connect all vertices of set $M$ to all vertices of set $N$, except for vertex $v_{M,m}$ that is connected only to $v_{N,n}$. However, each of the following pairs of vertices $\{v_{N,1}, v_{N,2}\}, \ldots, \{v_{N,n-3}, v_{N,n-2}\}$, together with vertex $v_{M,1}$ define one area bounded by two edges (between $v_{M,1}$ and the two vertices of set $N$) and a portion of the $y$-axis between the two vertices of set $N$. In each of these areas, we can place one auxiliary vertex: $v'_{M,m}, v''_{M,m}, v'''_{M,m}, \ldots$ Each of these auxiliary vertices can be connected to $v_{M,m}$ by a 3D edge, and to the respective pair of vertices of set $N$ that define the area in which the auxiliary vertex is placed by two planar edges. After this step, there will be only one missing edge between vertices $v_{M,m}$ and $v_{N,n-1}$, and we directly connect these two vertices by a single 3D edge, see Fig. 8(c).

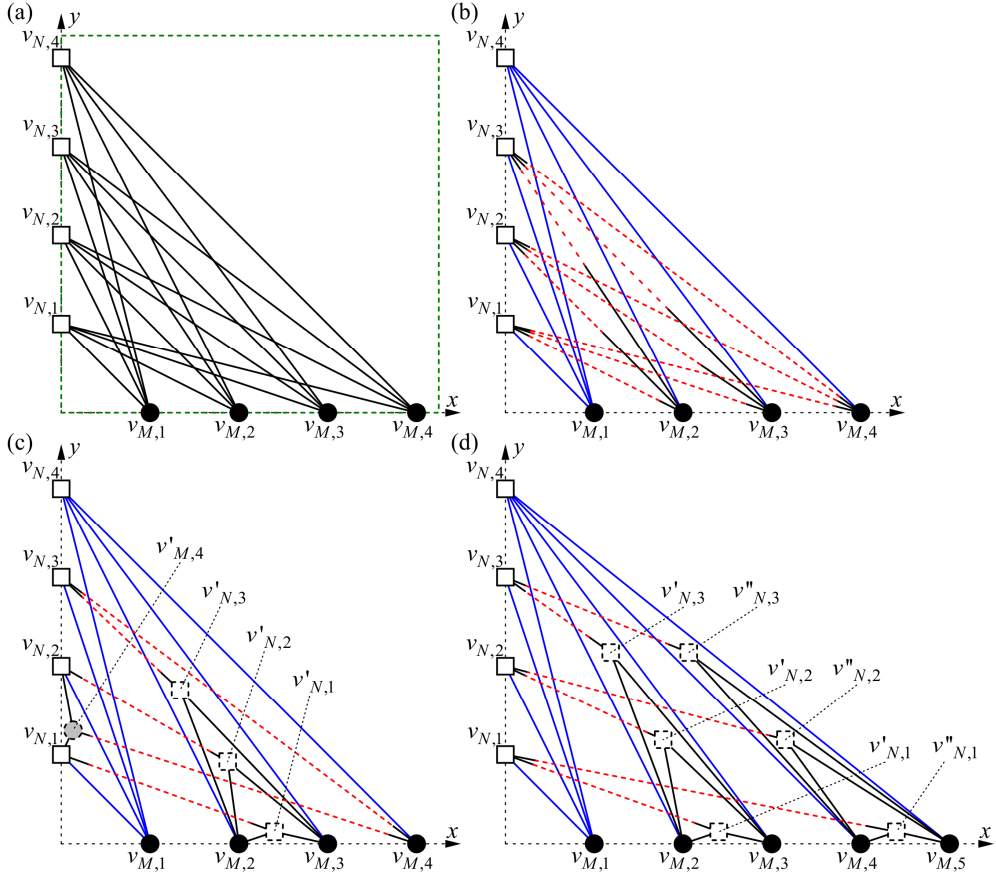

**Fig. 8.** Different graph drawings of a simplistic model of a facet-coupled $4 \times 4$ and $5 \times 4$ SAS circuit: **(a)** Graph drawing of a $4 \times 4$ SAS circuit where $1 \times 4$ and $4 \times 1$ switches at the input and output ports are LE. **(b)** Planar-edge-crossing-free graph representation of the same circuit. The edges depicted in blue represent a spanning planar subgraph of $K_{4,4}$. The remaining edges are realized with help of 3D edges (representing WOP) depicted as dashed red lines. The 3D edges connect to planar edges depicted in black that connect to the vertices in the drawing plane. **(c)** If the $1 \times 5$ and $5 \times 1$ switches at the input and output ports are realized as BT of $1 \times 2$ and $2 \times 1$ switches, the number of necessary 3D edges can be reduced by splitting the vertices of the original $K_{4,4}$ and by placing them into appropriate areas which are defined by the edges of the spanning planar subgraph (blue) and by the $x$ or $y$ coordinate axes. The white dashed squares represent $1 \times 2$ switches, while the filled gray circle represents a $2 \times 1$ switch. This approach allows to replace two 3D edges by one. In case where both $m$ and $n$ are even, the number of missing edges is odd, therefore, one missing edge (here: $\{v_{N,3}, v_{M,4}\}$) must be realized with help of one single 3D edge **(d)** Graph drawing of a $5 \times 4$ SAS circuit, analogous to the case described in (c). In case at least one of the numbers $m$ or $n$ is odd, the number of 3D edges can be reduced exactly 2 times compared to the case when $1 \times n$ and $m \times 1$ switches are realized as LE.

In case when at least one of the numbers $m$ and $n$ is odd, we can assume without loss of generality that $m$ is odd, and $n$ is even. By executing Step 1 as described above, we will get a subgraph where each of the pairs of vertices $\{v_{M,2}, v_{M,3}\}, \{v_{M,4}, v_{M,5}\}, \ldots, \{v_{M,m-1}, v_{M,m}\}$, together with vertex $v_{N,n}$ defines one area bounded by two edges (between $v_{N,n}$ and the two vertices of set $M$) and a portion of the $x$-axis between the two vertices of set $M$. After performing Step 2 as described above, we will connect all vertices of set $M$ to all vertices of set $N$. This case is illustrated on an example of $K_{5,4}$ in Fig. 8(d). For at least one of the numbers $m$ and $n$ being odd, the number of missing edges is even, and we can replace two missing planar edges by one 3D edge. Combining the two cases leads to Eq. (C5).

**Funding**

Erasmus Mundus Joint Doctorate program EUROPHOTONICS (Grant No. 159224-1-2009-1-FR-ERA MUNDUS-EMJD); Deutsche Forschungsgemeinschaft (DFG) – CRC 1173 (Wave Phenomena), project C4; European Research Council (ERC), Consolidator Grant 'TeraSHAPE', # 773248; BMBF project PRIMA (13N14630); H2020 Photonic Packaging Pilot Line 'PIXAPP' (# 731954); Alfried Krupp von Bohlen und Halbach Foundation; Karlsruhe School of Optics & Photonics (KSOP); Helmholtz International Research School for Teratronics (HIRST); Karlsruhe Nano-Micro Facility (KNMF).

**Acknowledgements**

We acknowledge fruitful discussions with Maria Axenovich (Institute of Algebra and Geometry (IAG), Karlsruhe Institute of Technology (KIT), Germany). Portions of this work were presented at the European Conference on Optical Communications ECOC in 2016 [21].